\documentclass[12pt]{article}  
\usepackage{epsfig}

\def\beq{\begin{equation}}
\def\eeq{\end{equation}}

\def\mpl{M_{\rm Pl}}

\def\eq#1{eq.~(\ref{#1})}
\def\fig#1{fig.~\ref{#1}}
\def\beqa{\begin{eqnarray}}
\def\eeqa{\end{eqnarray}}

\def\sps{Split Supersymmetry}
\def\mtil{\widetilde{m}}

\def\be{\begin{equation}}
\def\ee{\end{equation}}
\def\bc{\begin{center}}
\def\ec{\end{center}}
\def\bea{\begin{eqnarray}}
\def\eea{\end{eqnarray}}

\def\nn{\nonumber}
\def\ov{\overline}

\def\wt{\widetilde}
\def\sss{\scriptscriptstyle}
\def\PL{P_{L}}
\def\PR{P_{R}}

\def\mgl{m_{\tilde{g}}}

\def\mch{m_{\chi}}
\def\mcpi{m_{\chi^+_i}}
\def\mczi{m_{\chi^0_i}}
\def\sut{s_{13}}
\def\sdt{s_{23}}

\def\msusy{\wt{m}}
\def\QB{Q^{\,\wt{\scriptscriptstyle B}}}
\def\QG{Q^{\,\wt{\scriptscriptstyle G}}}
\def\QW{Q^{\,\wt{\scriptscriptstyle W}}}
\def\QH{Q^{\,\wt{\scriptscriptstyle H}}}
\def\CB{C^{\,\wt{\scriptscriptstyle B}}}
\def\CG{C^{\,\wt{\scriptscriptstyle G}}}
\def\CW{C^{\,\wt{\scriptscriptstyle W}}}
\def\CH{C^{\,\wt{\scriptscriptstyle H}}}
\def\QN{Q^{\,{\chi}^0_i}}
\def\QC{Q^{\,{\chi}^+_i}}
\def\CN{C^{\,{\chi}^0_i}}
\def\CC{C^{\,{\chi}^+_i}}
\def\CCL#1{\CC_{#1\,\sss L}}
\def\CCR#1{\CC_{#1\,\sss R}}
\def\CNL#1{\CN_{#1\,q_{\sss L}}}
\def\CNR#1{\CN_{#1\,q_{\sss R}}}
\def\sq2{\sqrt{2}}
\def\lle{_{\sss L}}
\def\rr{_{\sss R}}
\def\lr{_{\sss L,R}}
\def\rl{_{\sss R,L}}
\def\tg{\tau_{\tilde{g}}}
\def\gtot{\Gamma_{\rm tot}}
\def\as{\alpha_s}
\def\at{\alpha_t}

\catcode`@=11
\long\def\@caption#1[#2]#3{\par\addcontentsline{\csname
  ext@#1\endcsname}{#1}{\protect\numberline{\csname
  the#1\endcsname}{\ignorespaces #2}}\begingroup
    \small
    \@parboxrestore
    \@makecaption{\csname fnum@#1\endcsname}{\ignorespaces #3}\par
  \endgroup}
\catcode`@=12

\begin{document}

\baselineskip=18pt

\setcounter{footnote}{0}
\setcounter{figure}{0}
\setcounter{table}{0}

\begin{titlepage}
June 2005 \hspace*{\fill} CERN--TH/2005--106 %\hspace{0.6cm}
\newline  \hspace*{\fill} IPPP/05/25 DCTP/05/50
\newline  \hspace*{\fill} DFTT--17/05 %\hspace{1.2cm}

\begin{center}
\vspace{1cm}

{\Large \bf Gluino Decays in Split Supersymmetry}

\vspace{1cm}

{\large
P. Gambino$^{\,a}$,
G.F. Giudice$^{\,b}$,
P. Slavich$^{\,c}$
} 

\setcounter{footnote}{0}

\vspace{.5cm}

$^{ a}$ {\it INFN, Torino \& Dip.\ Fisica Teorica, Univ.\ di Torino,
I--10125 Torino, Italy}\\
$^{ b}$ {\it CERN, Theory Division, CH--1211 Geneva 23, Switzerland}\\
$^{ c}$ {\it Durham University, IPPP, DH1--3LE Durham, United Kingdom}

\end{center}
\vspace{1cm}

\begin{abstract}
\medskip
We compute the gluino lifetime and branching ratios in Split
Supersymmetry. Using an effective--theory approach, we resum the large
logarithmic corrections controlled by the strong gauge coupling and
the top Yukawa coupling. We find that the resummation of the radiative
corrections has a sizeable numerical impact on the gluino decay width
and branching ratios. Finally, we discuss the gluino decays into
gravitino, relevant in models with direct mediation of supersymmetry
breaking.

\end{abstract}

\bigskip
\bigskip

\end{titlepage}

%%%%%%%%%%%%%%%%%%%%%%%%%%%%%%%%%%%%%%%%%%%%%%%%%%%%%%%%%%%%%%%

\section{Introduction}
\label{sec1}

The long gluino lifetime is a trademark of \sps\
\cite{savas,split,noi4}.  The experimental discovery of a
slowly--decaying gluino~\cite{gluino} would not only be a strong
indication for {\sps}, but it would also allow for a measurement of
the effective supersymmetry--breaking scale $\mtil$, which cannot be
directly extracted from particle dynamics at the LHC. Moreover, the
gluino lifetime is a crucial parameter to determine the cosmological
constraints on the theory \cite{savas,arvanietal}. Therefore, for both
experimental and theoretical considerations, it is very important to
have a precise prediction of the gluino lifetime and branching ratios.

For what concerns the gluino decay processes in the MSSM, tree--level
results for the decays into chargino or neutralino and two quarks and
one--loop results for the radiative decay into neutralino and gluon
can be found in the literature \cite{decays}. In {\sps}, however, the
quantum corrections to the gluino decay processes can be very
significant, because they are enhanced by the potentially large
logarithm of the ratio between the gluino mass $\mgl$ and the scale
$\mtil$ at which the interactions responsible for gluino decay are
mediated. A fixed--order calculation of these processes in {\sps}
would miss terms that are enhanced by higher powers of the large
logarithm. In order to get a reliable prediction for the gluino decay
width, the large logarithmic corrections have to be resummed by means
of standard renormalization group techniques.

Recently, a calculation of the gluino decay widths in \sps\ was
presented in ref.~\cite{jim}, working at tree level for 3--body decays
and in (not resummed) one--loop approximation for 2--body decays.  In
this paper we will present a calculation of the gluino decay processes
that includes all leading corrections in $\as$ and $\at$, the strong
and top--Yukawa coupling constants.  As we will show, the inclusion
and resummation of leading--order corrections give sizeable
modifications of the gluino branching ratios, even for moderate values
of $\mtil$.

The structure of the paper is as follows: in sect.~\ref{sec2} we list
the operators in the effective Lagrangian of {\sps} that are
responsible for the decays of the gluino, and the high--energy
boundary conditions on the corresponding Wilson coefficients; in
sect.~\ref{sec3} we determine the renormalization group evolution of
the Wilson coefficients, and we express the operators in the
low--energy effective Lagrangian in terms of mass eigenstates; in
sect.~\ref{sec4} we discuss our numerical results for the branching
ratios and total width of the gluino decays in {\sps}; in
sect.~\ref{sec5} we consider the possibility of gluinos decaying into
gravitino; in sect.~\ref{concl} we present our conclusions. Finally,
in the appendix we provide the analytical formulae for the gluino
decay widths.

\section{The Effective Lagrangian}
\label{sec2}

Below the squark and slepton mass scale $\mtil$, the effective
Lagrangian of \sps\ describes the dynamics of Standard Model (SM)
particles together with higgsinos and gauginos. At the level of
renormalizable interactions, there is a conserved $G$--parity (under
which only the gluino is odd) preventing gluino decay. However,
integrating the squarks out of the underlying supersymmetric theory
induces non--renormalizable interactions that violate the $G$--parity.
Restricting our analysis up to dimension--6 operators, the $G$--odd
effective Lagrangian at the matching scale $\mtil$ is given by
\be
{\cal L} = \frac{1}{\msusy^2}\,\sum_{i=1}^7 \CB_i\,\QB_i \;+\;
\frac{1}{\msusy^2}\,\sum_{i=1}^2 \CW_i\,\QW_i \;+\; 
\frac{1}{\msusy^2}\,\left(\sum_{i=1}^5 \CH_i\,\QH_i + {\rm h.c.}\right) .
\ee

We are working in the basis of interaction eigenstates for gauginos
and higgsinos, neglecting the effect of electroweak symmetry breaking,
since $\mtil \gg M_Z$. The $G$--odd operators involving the $B$--ino
($\wt{B}$) are
\bea
\QB_1 & = &  \ov{\wt{B}} \,\gamma^{\mu}\,
\gamma_{5}\, {\tilde g}^a\; \otimes\;\sum_{k=1}^2 \;
\ov{q}\lle^{\,(k)} \,\gamma_{\mu} \, T^a \,q\lle^{\,(k)}
\label{qb1}\\
\QB_2 & = &  \ov{\wt{B}} \,\gamma^{\mu}\,
\gamma_{5}\, {\tilde g}^a\; \otimes\;
\sum_{k=1}^2\;\ov{u}\rr^{\,(k)} \,\gamma_{\mu} \, T^a \, u\rr^{\,(k)} \\
\QB_3 & = &  \ov{\wt{B}} \,\gamma^{\mu}\,
\gamma_{5}\, {\tilde g}^a\; \otimes\;
\sum_{k=1}^2\;\ov{d}\rr^{\,(k)} \,\gamma_{\mu} \, T^a\, d\rr^{\,(k)} \\
\QB_4 & = &  \ov{\wt{B}} \,\gamma^{\mu}\,
\gamma_{5}\, {\tilde g}^a\; \otimes\;
\ov{q}\lle^{\,(3)} \,\gamma_{\mu} \, T^a \,q\lle^{\,(3)}\\
\QB_5 & = &  \ov{\wt{B}} \,\gamma^{\mu}\,
\gamma_{5}\, {\tilde g}^a\; \otimes\;
\ov{t}\rr \,\gamma_{\mu} \, T^a \, t\rr \\
\QB_6 & = &  \ov{\wt{B}} \,\gamma^{\mu}\,
\gamma_{5}\, {\tilde g}^a\; \otimes\;
\ov{b}\rr \,\gamma_{\mu} \ T^a\, b\rr \\
\QB_7 & = & \ov{\wt{B}} \,\sigma^{\mu\nu} \,
\gamma_5 \, {\tilde g}^a\;G^a_{\mu\nu}  ,\label{qb7} 
\eea
where $k$ is a generation index, $T^a$ are the SU(3) generators and
$G^a_{\mu\nu}$ is the gluon field strength.  Assuming that the squark
mass matrices are flavour--diagonal, the Wilson coefficients of the
operators $\QB_i$ at the matching scale $\mtil$ are
\bea
\label{matchB1}
\CB_1(\mtil )=\CB_4(\mtil ) = - \frac{g_s\,g^{\prime}}{6}\,
r_{\tilde{q}\lle} \,, &&
\CB_2(\mtil )=\CB_5(\mtil ) =  \frac{2\,g_s\,g^{\prime}}{3}\, 
r_{\tilde{u}\rr}   \,,
\\
\label{matchB2}
\CB_3(\mtil )=\CB_6(\mtil ) = -  
\frac{g_s\,g^{\prime}}{3}\,r_{\tilde{d}\rr} \,, &&
\CB_7 (\mtil )= \frac{g_s^2\,g^{\prime}}{128\,\pi^2}\,(\mgl-m_{\sss\wt{B}})\,
\sum_{q} \,(r_{\tilde{q}\lle}-r_{\tilde{q}\rr})\,Q_q  \,,
\label{cb7}
\eea
where $r_{\tilde{q}} = \msusy^2/m_{\tilde{q}}^2$. Note that $\CB_7$
vanishes for mass--degenerate squarks.

The $G$--odd operators involving the $W$--ino ($\wt{W}$) are
\bea
\QW_1 & = & \ov{\wt{W}^{\sss A}} \,\gamma^{\mu}\,\gamma_5
\, {\tilde g}^a\; \otimes\;\sum_{k=1}^2 \;
\ov{q}\lle^{\,(k)} \,\gamma^{\mu} \,\tau^{\sss A}\,T^a\, q\lle^{\,(k)}
\label{qw1}\\
\QW_2 & = & \ov{\wt{W}^{\sss A}} \,\gamma^{\mu}\,\gamma_5
\, {\tilde g}^a\; \otimes\;
\ov{q}\lle^{\,(3)} \,\gamma^{\mu} \,\tau^{\sss A}\,T^a\, q\lle^{\,(3)} ,
\label{qw2}
\eea
where $\tau^{\sss A}$ are the Pauli matrices.  The matching conditions for
the Wilson coefficients are
\be
\label{matchW}
\CW_1(\mtil )=\CW_2(\mtil ) = - \frac{g_s\,g}{2}\, r_{\tilde{q}\lle} .
\ee

For the higgsinos, we use a compact notation in which the two Weyl
states $\wt{H}_u$ and $\wt{H}_d$ are combined in a single Dirac
fermion $\wt{H}\equiv \wt{H}_u + \varepsilon \,\wt{H}_d^c$, where
$\varepsilon$ is the antisymmetric matrix (with $\varepsilon_{12}=1$)
acting on the SU(2) indices. The states $\wt{H}_u$ and $\wt{H}_d$ can
be recovered by chiral decomposition, $\wt{H}_u =\wt{H}\lle$ and
$\wt{H}_d =-\varepsilon \,(\wt{H}^c)\lle$. Keeping only the
third--generation Yukawa couplings, the $G$--odd operators involving
higgsinos are
\bea
\QH_1 & = &  \ov{\wt{H}}\lle \, {\tilde g}^a\rr \; \otimes\;
\varepsilon \, \ov{q}\lle^{\,(3)}  \, T^a\, t\rr 
\label{qh1}\\
\QH_2 & = &  \ov{\wt{H}}\lle \,\sigma^{\mu\nu} \,
 {\tilde g}^a\rr\; \otimes\; 
\varepsilon \,\ov{q}\lle^{\,(3)} \,\sigma_{\mu\nu}  \, T^a\,t\rr \\
\QH_3 & = &  \ov{\wt{H}}\rr \, {\tilde g}^a\lle\; \otimes\;
\ov{b}\rr  \, T^a\,q\lle^{\,(3)} \\
\QH_4 & = &  \ov{\wt{H}}\rr \,\sigma^{\mu\nu} \,
 {\tilde g}^a\lle \; \otimes\; 
\ov{b}\rr \,\sigma_{\mu\nu}  \,T^a\, q\lle^{\,(3)}\\
\QH_5 & = & \ov{\wt{H}}\lle \,\sigma^{\mu\nu} \,
 {\tilde g}^a\rr\; h\;G^a_{\mu\nu},
\label{qh5}
\eea
where $h$ is the Higgs doublet. The Wilson coefficients at the
matching scale $\mtil$ are
\be
\label{matchH1}
\CH_1(\mtil ) = \frac{g_s\,h_t}{\sq2\sin\beta}\,
(r_{\tilde{q}\lle}-r_{\tilde{u}\rr})\,,\;\;\;\;\;\;
\CH_2(\mtil ) = \frac{g_s\,h_t}{4\,\sq2\sin\beta}\,
(r_{\tilde{q}\lle}+r_{\tilde{u}\rr})\,,
\ee
\be
\label{matchH2}
\CH_3(\mtil ) = \frac{g_s\,h_b}{\sq2\cos\beta}\,
(r_{\tilde{q}\lle}-r_{\tilde{d}\rr})\,,\;\;\;\;\;\;
\CH_4(\mtil ) = -\frac{g_s\,h_b}{4\,\sq2\cos\beta}\,
(r_{\tilde{q}\lle}+r_{\tilde{d}\rr})\,,
\ee
\be
\CH_5(\mtil ) = \frac{g_s^2\,h_t^2}{32\,\sq2\,\pi^2\,\sin\beta}
(r_{\tilde{q}\lle}+r_{\tilde{u}\rr}) .
\label{ch5}
\ee
Here $h_t$ and $h_b$ are the top and bottom Yukawa couplings,
and $\tan\beta$ is a free parameter of {\sps}.

Before proceeding to the operator renormalization, we want to make
some remarks.

{\it (i)} We recall that all coupling constants appearing in the
expressions of the Wilson coefficients given above have to be computed
at the scale $\mtil$.

{\it (ii)} Note that we have given the Wilson coefficients of the
4--fermion operators at the leading perturbative order, while the
coefficients of the operators $\QB_7$ and $\QH_5$ are given at the
next order (one--loop approximation). The operator anomalous
dimensions will be computed in sect.~\ref{sec3} at the leading order
in the strong and top--Yukawa couplings $\as=g_s^2/(4\pi)$ and $\at
=h_t^2/(4\pi)$.  Therefore, the gluino 3--body decays, mediated only
by 4--fermion operators, will be computed by resumming all
$\alpha_{s,t}\ln (\mtil / \mgl )$ corrections, but neglecting terms
${\cal O} [\alpha^{n+1}_{s,t}\ln^n (\mtil / \mgl )]$ with $n \ge
0$. For the radiative 2--body gluino decay into a gluon and a
neutralino, a greater accuracy is more appropriate. The expressions of
$\CB_7$ and $\CH_5$ given in eqs.~(\ref{cb7}) and (\ref{ch5}),
together with leading--order anomalous dimensions and one--loop matrix
elements [see eq.~(\ref{coeffg}) below], allow us to determine the
2--body decay amplitude neglecting terms ${\cal O}
[\alpha^{n+1}_{s,t}\ln^n (\mtil / \mgl )]$ with $n \ge 1$. This means
that we have resummed all large logarithms at the leading order in all
cases, but our formulae for 2--body gluino decays contain also the
complete ${\cal O} (\alpha_{s,t})$ terms, relevant when the logarithm
is not large.

{\it (iii)} If $\mtil$ is close to the GUT scale, in presence of
gauge--coupling unification there is no solid justification for the
approximation of computing $\as$ contributions to the anomalous
dimensions, neglecting electroweak corrections. However, because of
the large SU(3) coefficients, we consider our approximation to be
fairly adequate, even for $\mtil$ as large as $10^{13}$~GeV, which is
the maximum value of $\mtil$ consistent with the negative searches for
anomalous heavy isotopes.

{\it (iv)} In \eq{matchH2} we have included the contribution from the
bottom Yukawa coupling $h_b$, since these coefficients are enhanced when
$\tan\beta$ is large. There are no $\tan\beta$ enhancements in the
evolution below $\mtil$, and therefore our results are reliable for any
value of $\tan\beta$.

{\it (v)} \sps\ is free from flavour problems, therefore our
assumption that squark mass matrices are diagonal is unnecessary.  On
the other hand, a certain degree of mass degeneracy among squarks is
required by gauge-coupling unification. In the results presented in
sect.~\ref{sec4} we take for simplicity all squark masses to be equal.

\section{Operator Renormalization}
\label{sec3}

The renormalization--group flow for the Wilson coefficients is
determined by the equations
\bea
\mu\frac{d\vec C}{d \mu}&=&  {\hat \gamma}^T(\as,\at) \,\vec C 
\label{runc}\\
\mu\frac{d\as}{d \mu}&=& -\beta_s \,\frac{\as^2}{2\pi}\\
\mu\frac{d\at}{d \mu}&=& -\beta_t \,\frac{\at^2}{2\pi}-\beta_{st}
 \frac{\as \at}{2\pi},
\eea
where $\mu$ is the renormalization scale and,
in \sps, we have $\beta_s=5$, $\beta_t=-9/2$ and $\beta_{st}=8$.
The anomalous--dimension matrix $\hat \gamma$ can be expressed as
\beq
{\hat \gamma}_{ij}=-2\,b_{ij}-\delta_{ij}\sum_f a_f ,
\label{gaman}
\eeq
where $b_{ij}$ are extracted from the poles of the one--loop
renormalization of the operators $Q_i$ ($Q_i \rightarrow b_{ij}\,
Q_j/\epsilon + \cdots$). In \eq{gaman} the sum is over all fields
entering the operator $Q_i$, and the field anomalous dimensions $a_f$
are given by
\be
\label{wfrs}
a_{q^k\lle} = -\frac{1}{4\,\pi} \left(
\as \,C_F + \frac{\at}{2}\,\delta_{k3}\right)
\,,\;\;\;\;\;
a_{u^k\rr}= -\frac{1}{4\,\pi} \left(
\as \,C_F + \at\,\delta_{k3}\right)
\,,\;\;\;\;\;
a_{d\rr} = -\frac{\as \,C_F}{4\,\pi}\,,
\ee
\be
a_{\tilde{g}} = -\frac{\as \,N_c}{4\,\pi}
\,,\;\;\;\;\;
a_{h} = -\frac{\at \,N_c}{4\,\pi}
\,,\;\;\;\;\;
a_g = \frac{\as}{4\,\pi} \; \left( N_c - \frac 23 \,N_f \right) .
\ee
Here $k$ is a generation index, $C_F=(N_c^2-1)/(2N_c)$, $N_c=3$,
$N_f=6$. Note that the gluon anomalous dimension $a_g$ (given here in
the Feynman gauge) is different from the SM value because it includes
the gluino contribution.

We find that the anomalous--dimension matrices of the $B$--ino
operators in eqs.~(\ref{qb1})--(\ref{qb7}), of the $W$--ino operators
in eqs.~(\ref{qw1})--(\ref{qw2}), and of the higgsino operators in
eqs.~(\ref{qh1})--(\ref{qh5}) are respectively
\beq
{\hat \gamma}^{(a)}=\frac{\as}{4\pi} \gamma_s^a+
\frac{\at}{4\pi} \gamma_t^a+\frac{\sqrt{\as \at}}{4\pi} 
\gamma_{st}^a
,~~~~a=\wt{\sss B},\wt{\sss W},\wt{\sss H}
\label{gamgam}
\eeq

\beq
\gamma_s^{\wt{\sss B}}=\frac 13 \pmatrix{
8-9N_c & 8 & 8 & 8 & 8 & 8 & 0 \cr
4 & 4-9N_c & 4 & 4 & 4 & 4 & 0 \cr
4 & 4 & 4-9N_c & 4 & 4 & 4 & 0 \cr
4 & 4 & 4 & 4-9N_c & 4 & 4 & 0 \cr
2 & 2 & 2 & 2 & 2-9N_c & 2 & 0 \cr
2 & 2 & 2 & 2 & 2 & 2-9N_c & 0 \cr
0 & 0 & 0 & 0 & 0 &  0 & 2N_f-18 N_c },
\label{gammabino}
\eeq
\beq
\gamma_t^{\wt{\sss B}}= \pmatrix{
0 & 0 & 0 & 0 & 0 &  0 & 0 \cr
0 & 0 & 0 & 0 & 0 &  0 & 0 \cr
0 & 0 & 0 & 0 & 0 &  0 & 0 \cr
0 & 0 & 0 & 1 & \!\!\!\!-2 &  0 & 0 \cr
0 & 0 & 0 & \!\!\!\!-1 & 2 &  0 & 0 \cr
0 & 0 & 0 & 0 & 0 &  0 & 0 \cr
0 & 0 & 0 & 0 & 0 &  0 & 0 },~~~~\gamma_{st}^{\wt{\sss B}}=0,
\eeq

\beq
\gamma_s^{\wt{\sss W}}=\pmatrix{
-3N_c & 0 \cr
0 & -3N_c } ,~~~~
\gamma_t^{\wt{\sss W}}=\pmatrix{
0 & 0 \cr
0 & 1 },~~~~\gamma_{st}^{\wt{\sss W}}=0,
\eeq

\beq
\gamma_s^{\wt{\sss H}}=\pmatrix{
\frac{3}{N_c} & 0 & 0 & 0 & 0 \cr
0 & -4N_c-\frac{1}{N_c} & 0 & 0 & 0 \cr
0 & 0 & \frac{3}{N_c} & 0 & 0 \cr
0 & 0 & 0 & -4N_c-\frac{1}{N_c} & 0 \cr
0 & 0 & 0 & 0 & \frac 23 N_f -6N_c },
\eeq

\beq
\gamma_t^{\wt{\sss H}}=\frac 12 \pmatrix{
3 & 0 & 0 &  0 & 0 \cr
0 & 3 & 0 &  0 & 0 \cr
0 & 0 & 1 &  0 & 0 \cr
0 & 0 & 0 &  1 & 0 \cr
0 & 0 & 0 &  0 & 2N_c },~~~~
\gamma_{st}^{\wt{\sss H}}=\pmatrix{
0 & 0 & 0 &  0 & 0 \cr
0 & 0 & 0 &  0 & 4 \cr
0 & 0 & 0 &  0 & 0 \cr
0 & 0 & 0 &  0 & 0 \cr
0 & 2 & 0 &  0 & 0 } .
\eeq

\begin{figure}[t]
\begin{center}
\mbox{\epsfig{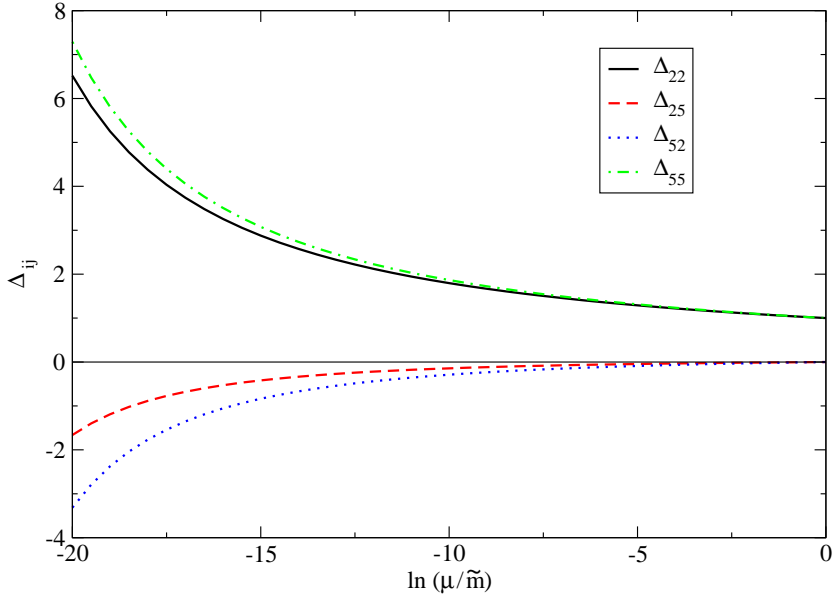}}
\end{center}
\caption{\sf Renormalization group flow of $\CH_2$ and $\CH_5$,
expressed in terms of the coefficients $\Delta_{ij}$ of \eq{Deltas},
for $\as(\mtil )=0.05$ and $\at(\mtil )=0.03$.  The solid, dashed,
dotted, and dot--dashed lines correspond to $\Delta_{22}$,
$\Delta_{25}$, $\Delta_{52}$ and $\Delta_{55}$, respectively.  }
\label{fig:coeffs}
\end{figure}

For coefficients with only multiplicative renormalization (which is
the case for $\CB_7$, $\CW_{1,2}$, $\CH_{1,3,4}$), \eq{runc} can be
easily integrated, with the result
\be
C_i(\mu)= C_i(\mtil ) \,\eta_s^{\left
(\frac{\gamma_s}{2\beta_s}-\frac{\beta_{st}
\gamma_t}{2 \beta_s\beta_t}\right)} \ 
\eta_t ^{\frac{\gamma_t}{2\beta_t}}~~~~~{\rm for~~}
C_i=\CB_7,~\CW_{1,2},~\CH_{1,3,4}  .
\ee
We have defined
\bea
\label{etas}
\eta_s \equiv \frac{\as(\mtil )}{\as(\mu)}&=&1+\frac{\as(\mtil )}{2\pi}
\beta_s \ln \frac{\mu}{\mtil} \,, \\
\label{etat}
\eta_t \equiv \frac{\at(\mtil )}{\at(\mu )}&=&
\eta_s^\frac{\beta_{st}}{\beta_s} + \frac{\at(\mtil )
\beta_t}{\as(\mtil ) \left( \beta_{st} -\beta_s \right)}
\left(  \eta_s^\frac{\beta_{st}}{\beta_s} -\eta_s \right) .
\eea
The evolution of the Wilson coefficients for the other $B$--ino
operators involves operator mixing and the solution of \eq{runc} is
given by
\bea
\CB_i (\mu) &=& \eta_s^{-\frac {9}{10}} \left[ \CB_i (\mtil) 
+y \,{\ov C}(\mtil )\right] 
~~~~i=1,2,3,6 \label{pinc}\\
\CB_4 (\mu) &=& \eta_s^{-\frac {9}{10}} \left[ (1+z)\,\CB_4 (\mtil)
-z \,\CB_5 (\mtil ) +y \,{\ov C}(\mtil )\right] ,\\
\CB_5 (\mu ) &=& \eta_s^{-\frac {9}{10}} \left[ (1+2z)\,\CB_5 (\mtil)
-2z \,\CB_4 (\mtil ) +y \,{\ov C}(\mtil )\right] ,\label{pall}
\eea
where ${\ov C}=\CB_1 /3 +(\CB_2+\CB_3+\CB_4)/6+(\CB_5+\CB_6)/12$,
$y=\eta_s^{\,4/5}-1$, and $z=(\eta_s^{\,8/15}\,\eta_t^{-1/3}-1)/3$.
Because of the non--vanishing contribution from $\gamma^{\wt{\sss
H}}_{st}$, the equations for $\CH_2$ and $\CH_5$ cannot be solved
analytically. The numerical results for the renormalization
coefficients $\Delta_{ij}$, defined by
\beq
\label{Deltas}
\pmatrix{\CH_2 (\mu)\cr \CH_5 (\mu)} =\pmatrix{
\Delta_{22} &\Delta_{25} \cr  \Delta_{52} &\Delta_{55}}
\pmatrix{\CH_2 (\mtil )\cr \CH_5 (\mtil )}\,,
\eeq
are shown in \fig{fig:coeffs} for a representative choice of $\as(\mtil )$
and $\at(\mtil )$. Despite the fact that the high--energy boundary
condition on $\CH_5$, eq.~(\ref{ch5}), is suppressed by a loop factor,
a sizeable value of $\CH_5$ can be generated through the mixing with
$\CH_2$.

A computation of the ${\cal O}(\as)$ part of the anomalous dimensions,
restricted to the four--fermion operators, has been given in the
appendix of ref.~\cite{arvanietal}. From the comparison with
eq.~(\ref{gammabino}) it appears that the authors of ref.~\cite{arvanietal}
have omitted the mixing among the $B$--ino operators induced by the
penguin diagrams. Also, we disagree with ref.~\cite{arvanietal} on the
anomalous dimensions of the higgsino operators.

Once we have evolved the Wilson coefficients down to the
renormalization scale at which we compute the gluino decay width, it
is convenient to express the operators in terms of chargino and
neutralino mass eigenstates.  With the usual definitions for the
chargino and neutralino mixing matrices $U$, $V$ and $N$, which we
assume to be real, the $B$--ino, $W$--ino and higgsino spinors can be
expressed as
\be
\ov{\wt{W}^+} = \ov{\chi^+_i}\,
\left(U_{i1}\,\PL+V_{i1}\,\PR\right),\;\;\;\;\;\;
\ov{\wt{H}^+} = \ov{\chi^+_i}\,
\left(U_{i2}\,\PL+V_{i2}\,\PR\right),
\ee
\be
\ov{\wt{B}} = \ov{\chi^0_i}\,N_{i1}\,,\;\;\;
\ov{\wt{W}^3} = \ov{\chi^0_i}\,N_{i2} \,,\;\;\;
\ov{\wt{H}^0} =  \ov{\chi^0_i}\,
\left(N_{i4}\, \PR - N_{i3}\,\PL\right) ,
\ee
where $\PL$ and $\PR$ are the chiral projectors. In the basis of mass
eigenstates, the effective Lagrangian becomes
\be \label{lagr2}
{\cal L} = 
\frac{1}{\msusy^2}\,\sum_{j} \CN_j\,\QN_j \;+\; 
\frac{1}{\msusy^2}\,\left(\sum_j \CC_j\,\QC_j + {\rm h.c.}\right) .
\ee

The operators involving neutralinos and quarks and their corresponding Wilson
coefficients are
\bea
\QN_{1\,q\lle,q\rr} &=& 
\ov{\chi^0_i}\,\gamma^{\mu}\,\gamma_5\,\,{\tilde g}^a\;\otimes\;
\sum_{k=1}^2\,\ov{q}^{\,(k)}\lr\,\gamma_{\mu}\,T^a\,q^{(k)}\lr
\hspace{2cm}(q=u,d)\,,
\\
\QN_{2\,q\lle,q\rr} &=& 
\ov{\chi^0_i}\,\gamma^{\mu}\,\gamma_5\,{\tilde g}^a\;\otimes\;
\ov{q}\lr\,\gamma_{\mu}\,T^a\,q\lr
\hspace{2.75cm}(q=t,b)\,,
\\
\QN_{3\,q\lle,q\rr} &=& 
\ov{\chi^0_i}\rl\,{\tilde g}^a\lr\;\otimes\;
\ov{q}\rl\,T^a\,q\lr
\hspace{3.45cm}(q=t,b)\,, 
\\
\QN_{4\,q\lle,q\rr} &=&
\ov{\chi^0_i}\rl\,\sigma^{\mu\nu}\,\gamma_5\, {\tilde g}^a\lr\;\otimes\;
\ov{q}\rl\,\sigma_{\mu\nu}\,T^a\,q\lr
\hspace{1.65cm}(q=t,b)\,,
\eea
\bea
\label{wils1}
&&
\CN_{1\,u\lle} = \CB_1\,N_{i1} + \CW_1\,N_{i2}\,,\;\;\;\;\;
\CN_{1\,u\rr} = \CB_2\,N_{i1}\,,\\
&& 
\CN_{1\,d\lle} = \CB_1\,N_{i1} - \CW_1\,N_{i2}\,,\;\;\;\;\;
\CN_{1\,d\rr} = \CB_3\,N_{i1}\,,\\
&&
\CN_{2\,t\lle} = \CB_4\,N_{i1} + \CW_2\,N_{i2}\,,\;\;\;\;
\CN_{3\,t\lle} = - \CH_1\,N_{i4}\,,\;\;\;\;\;
\CN_{4\,t\lle} = \CH_2\,N_{i4}\,,\\
&&
\CN_{2\,t\rr} = \CB_5\,N_{i1}\,,\;\;\;\;\;
\CN_{3\,t\rr} = - \CH_1\,N_{i4}\,,\;\;\;\;\;
\CN_{4\,t\rr} = -\CH_2\,N_{i4}\,,\\
&& 
\CN_{2\,b\lle} = \CB_4\,N_{i1} - \CW_2\,N_{i2}\;\;\;\;
\CN_{3\,b\lle} = - \CH_3\,N_{i3}\,,\;\;\;\;\;
\CN_{4\,b\lle} = -\CH_4\,N_{i3}\,,\\
&& 
\label{wils2}
\CN_{2\,b\rr} = \CB_6\,N_{i1}\,,\;\;\;\;\;
\CN_{3\,b\rr} = - \CH_3\,N_{i3}\,,\;\;\;\;\;
\CN_{4\,b\rr} = \CH_4\,N_{i3}.
\eea

The operators involving charginos and quarks and their Wilson
coefficients are
\bea
\QC_{1\,{\sss L,R}} &=& 
\ov{\chi^+_i}\lr\,\gamma^{\mu}\,{\tilde g}^a\lr\;\otimes\;
\sum_{k=1}^2 \,\ov{d}\lle^{\,(k)}\,\gamma_{\mu}\,T^a\,
u\lle^{(k)} \\
\QC_{2\,{\sss L,R}} &=&
\ov{\chi^+_i}\lr\,\gamma^{\mu}\,{\tilde g}^a\lr\;\otimes\;
\ov{b}\lle\,\gamma_{\mu}\,T^a\,t\lle \\
\QC_{3\,{\sss L,R}} &=&
\ov{\chi^+_i}\rl\,{\tilde g}^a\lr\;\otimes\;
\ov{b}\rl\,T^a\,t\lr \\
\QC_{4\,{\sss L,R}} &=&
\ov{\chi^+_i}\rl\,\sigma^{\mu\nu}\,{\tilde g}^a\lr\;\otimes\;
\ov{b}\rl\,\sigma_{\mu\nu}\,T^a\,t\lr
\eea
\bea
&&
\label{wils3}
\CCL{1} \;=\; -\sqrt{2}\,\CW_1\,V_{i1}\,,\;\;\;\;\;
\CCR{1} \;=\; \sqrt{2}\,\CW_1\,U_{i1}\,,
\\ &&
\CCL{2} = -\sqrt{2}\,\CW_2\,V_{i1}\,,\;\;\:
\CCL{3} = \CH_3\,U_{i2}\,,\;\;\;\;\;
\CCL{4} = \CH_4\,U_{i2}\,,\\
&&
\CCR{2} = \sqrt{2}\,\CW_2\,U_{i1}\,,\;\;\;\;\;
\CCR{3} = \CH_1\,V_{i2}\,,\;\;\;\;\;
\CCR{4} = \CH_2\,V_{i2}\,.
\label{wils4}
\eea 
All Wilson coefficients in eqs.(\ref{wils1})--(\ref{wils2}) and
(\ref{wils3})--(\ref{wils4}) are evaluated at the 
scale $\mu$ at which the gluino decay width is computed (we
take $\mu=\mgl$ in our numerical analysis).

The magnetic operator involving a neutralino and a gluon is
\be 
\QN_g =
\ov{\chi^0_i}\,\sigma^{\mu\nu}\,\gamma_5\,{\tilde g}^a\,G_{\mu\nu}^a\,.
\ee
In order to reach the desired accuracy in the ${\tilde g}\to g {\tilde
\chi}^0$ process, we need to include the matrix element contribution
coming from the diagram in which the two top quarks in the operator
$\QH_2$ close in a loop emitting a gluon. This results into an
``effective'' Wilson coefficient
\be
\label{coeffg}
\left.C^{\,{\chi}^0_i}_g\right._{\rm \!\!\!eff}
(\mu) =\CB_7(\mu)\,N_{i1}+\CH_5(\mu)\,N_{i4}\,v +
\frac{g_s\,h_t}{8\pi^2}\,\CH_2(\mu)\,\,N_{i4}\,v\,\ln\frac{m_t^2}{\mu^2}\,,
\ee
where $v$ is the Higgs vacuum expectation value and we
take $\mu =\mgl$.

From the effective Lagrangian of \eq{lagr2} we can compute the gluino
decay widths and complete expressions can be found in the appendix.
The same effective Lagrangian correctly describes also the
interactions that lead to the decays ${\tilde g}\to g g{\tilde
\chi}^0$ and ${\tilde g}\to g h^0 {\tilde \chi}^0$. However, since
these processes are subleading, we will not explicitly calculate their
decay widths.

\section{Results}
\label{sec4}

We are now ready to discuss the results of our computation of the
decay width and branching ratios of the gluino in {\sps}.  The input
parameters relevant to our analysis are the sfermion mass scale
$\mtil$, the physical gluino mass $\mgl$ and $\tan \beta$, which in
{\sps} is interpreted as the tangent of the angle that rotates the
finely tuned Higgs doublets. To simplify the analysis we assume that
the squark masses are degenerate, i.e.~we set $r_{\tilde{q}\lle} =
r_{\tilde{u}\rr} = r_{\tilde{d}\rr} = 1$ in the matching conditions of
the Wilson coefficients.  The gluino mass parameter in the Lagrangian,
$M_3$, is extracted from $\mgl$ including radiative corrections, and
the other gaugino masses $M_1$ and $M_2$ are computed from $M_3$
assuming unification at the GUT scale. The higgsino mass parameter
$\mu$ is determined as a function of $M_2$ by requiring that the relic
abundance of neutralinos is equal to the dark--matter density
preferred by WMAP data \cite{wmap} (see fig.~11 of
ref.~\cite{split}\,). The sign of $\mu$ remains a free parameter, but
since it does not affect our results for the gluino decays in a
significant way we will assume $\mu>0$ throughout our analysis. The
effective couplings of gauginos and higgsinos at the weak scale,
needed to compute the chargino and neutralino mass matrices, are
determined from their high--energy (supersymmetric) boundary values by
means of the renormalization--group equations of {\sps}, given in
ref.~\cite{split}.  Finally, the SM input parameters relevant to our
analysis are: the physical masses for the top quark and gauge bosons,
$m_t=178$ GeV, $M_Z = 91.187$ GeV and $M_W=80.41$ GeV; the running
bottom mass computed at the scale of the top mass, $m_b(m_t) = 2.75$
GeV; the Fermi constant, $G_F = 1.166 \times 10^{-5}$ GeV$^{-2}$; the
running strong coupling computed at the scale of the top mass,
$\as(m_t) = 0.106$.

\begin{figure}[t]
\begin{center}
\mbox{\epsfig{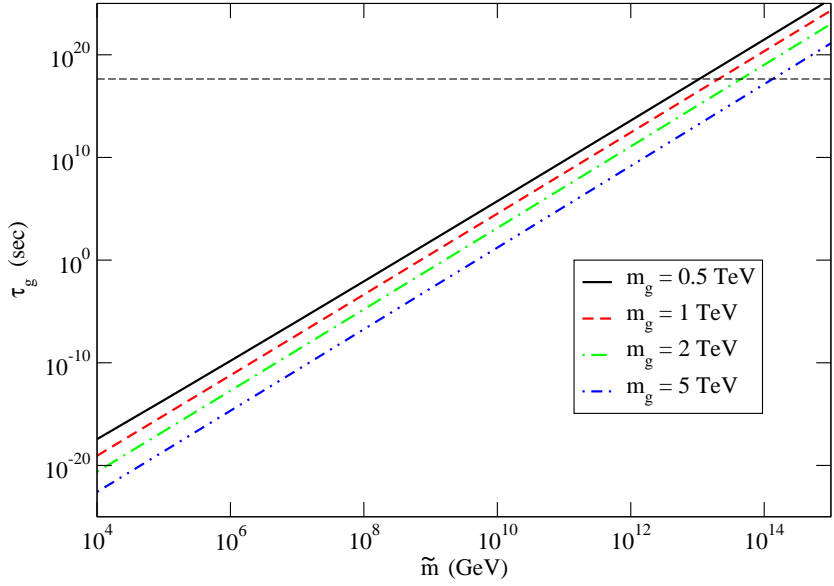}}
\end{center}
\vspace{-2mm}
\caption{\sf Gluino lifetime $\tg$ as a function of the sfermion mass
scale $\msusy$, for different values of the physical gluino mass
$\mgl$. The other free parameters are chosen as $\tan\beta = 2$ and
$\mu>0$. The dashed horizontal line corresponds to the age of the universe,
$\tau_{\sss U}=14$ Gyr.}
\label{fig:life}
\end{figure}

To start our discussion, we show in \fig{fig:life} the gluino lifetime
$\tg$ (in seconds) as a function of the sfermion mass scale $\msusy$,
for $\tan\beta = 2$ and four different values of the physical gluino
mass ($\mgl$ = 0.5, 1, 2 and 5 TeV, respectively).  It can be seen
that $\tg$ is about 4 seconds for $\mgl=1$ TeV and $\msusy = 10^9$
GeV. A value of $\tg$ equal to the age of the universe (14~Gyr)
corresponds to $\mtil = (1.1, \, 2.1,\, 4.5,\, 13) \times 10^{13}$ GeV
for $\mgl$ = 0.5, 1, 2 and 5 TeV, respectively.

\begin{figure}[p]
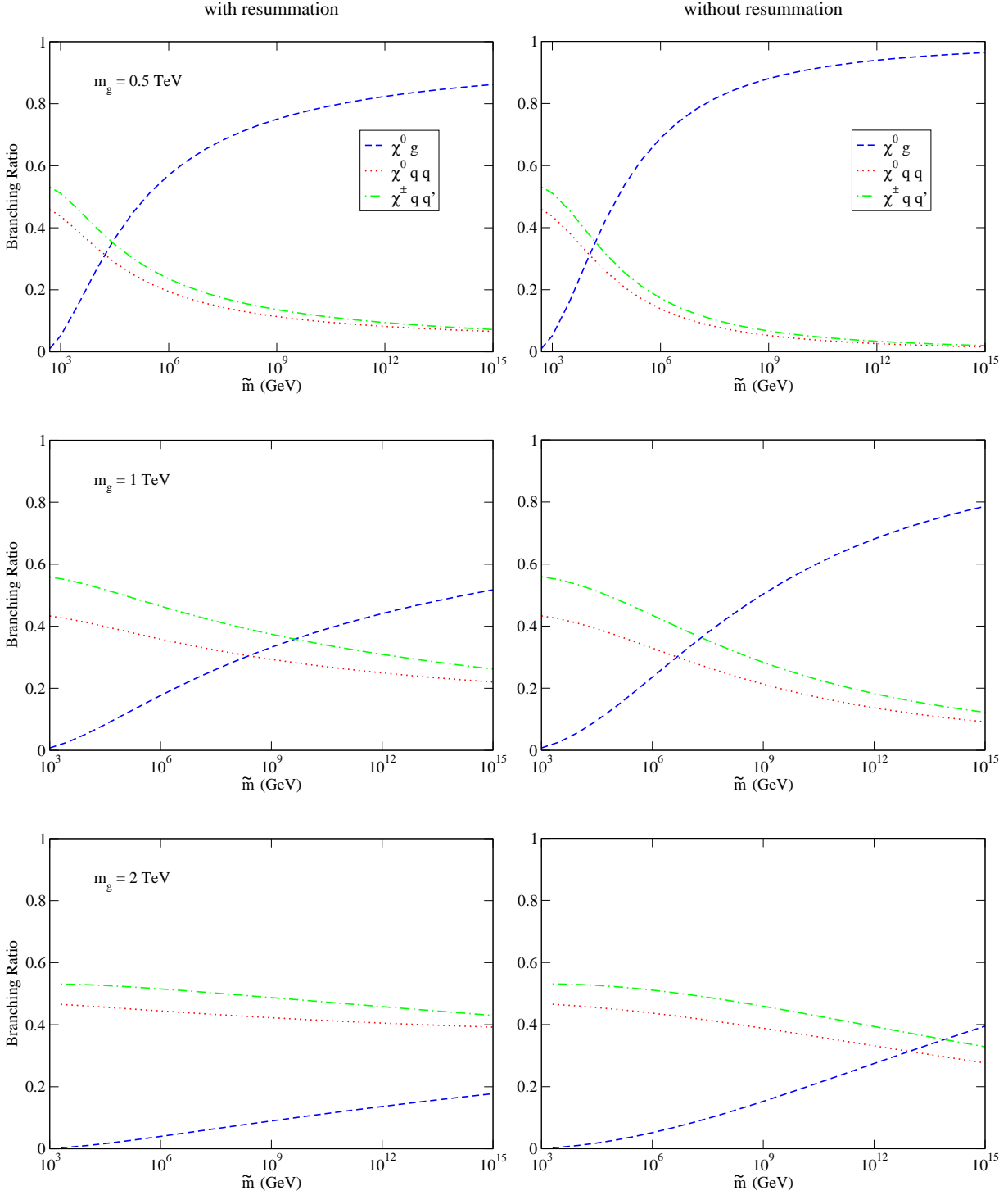

\begin{center}

\mbox{
\epsfig{figure=BRvsMS_tb20_mg500_plus.eps,width=8.15cm}
\epsfig{figure=BRvsMS0_tb20_mg500_plus.eps,width=7.85cm}}

\vspace*{6mm}

\mbox{
\epsfig{figure=BRvsMS_tb20_mg1000_plus.eps,width=8.15cm}
\epsfig{figure=BRvsMS0_tb20_mg1000_plus.eps,width=7.85cm}}

\vspace*{6mm}

\mbox{
\epsfig{figure=BRvsMS_tb20_mg2000_plus.eps,width=8.15cm}
\epsfig{figure=BRvsMS0_tb20_mg2000_plus.eps,width=7.85cm}}

\end{center}
\caption{\sf Branching ratios for the gluino decay channels $\chi^0 g$
(dashed lines), $\chi^0 q \bar{q}$ (dotted) and $\chi^{\pm} q
\bar{q}^{\,\prime}$ (dot--dashed), summed over all possible neutralino
or chargino states, as a function of $\msusy$, for three values of
$\mgl$: 500 GeV (upper plots), 1 TeV (middle) and 2 TeV (lower).  The
curves in the left (right) plots do (do not) include the resummation
of the leading logarithmic corrections. Other parameters are
$\tan\beta=20$ and $\mu>0$.}
\label{fig:bratios}
\end{figure}

In \fig{fig:bratios} we show the branching ratios for the three decay
processes $\tilde{g} \rightarrow \chi^0 g$, $\tilde{g} \rightarrow
\chi^0 q \bar{q}$ and $\tilde{g} \rightarrow \chi^{\pm} q
\bar{q}^{\,\prime}$ (summed over all neutralino or chargino states) as
a function of $\msusy$, for $\tan\beta = 20$ and three different
values of $\mgl$\,: 500 GeV (upper plots), 1 TeV (middle plots) and 2
TeV (lower plots). The value of $\tan\beta$ has little impact on these
results.  The plots on the left of \fig{fig:bratios} represent the
results of our full calculation, including the resummation of the
leading logarithmic corrections controlled by $\as$ and $\at$. The
plots on the right represent instead the lowest--order results that do
not include the resummation.  We obtain the latter results by
replacing the Wilson coefficients of the four--fermion operators in
the low--energy effective Lagrangian with their tree--level
expressions in terms of gauge and Yukawa couplings
[eqs.~(\ref{matchB1})--(\ref{matchB2}), (\ref{matchW}) and
(\ref{matchH1})--(\ref{matchH2})], and the Wilson coefficient of the
magnetic operator with its one--loop expression.
The plots in \fig{fig:bratios} show that the branching ratio of the
radiative decay $\tilde{g} \rightarrow \chi^0 g$ decreases for
increasing $\mgl$ and increases for increasing $\msusy$. In fact, as
stressed in ref.~\cite{jim}, the ratio between the two--body and
three--body decay rates computed at lowest order scales like
$m_t^2/\mgl^2\;[1-\ln(\msusy^2/m_t^2)]^2$, where the logarithmic term
comes from the top--stop loop that generates the magnetic
gluino--gluon--higgsino interaction. For large values of $\msusy$, the
resummation of the logarithms becomes necessary.  Comparing the plots
on the left and right sides of \fig{fig:bratios}, we see that the
resummation of the leading logarithmic corrections tends to enhance
the three--body decays and suppress the radiative decay. The effect of
the corrections on the branching ratios is particularly visible when,
like in the middle and lower plots, neither the two--body nor the
three--body channels are obviously dominant in the range $10^8$ GeV $<
\msusy < 10^{13}$ GeV, relevant to \sps.

\begin{figure}[t]
\begin{center}
\mbox{\hspace{8mm}
\epsfig{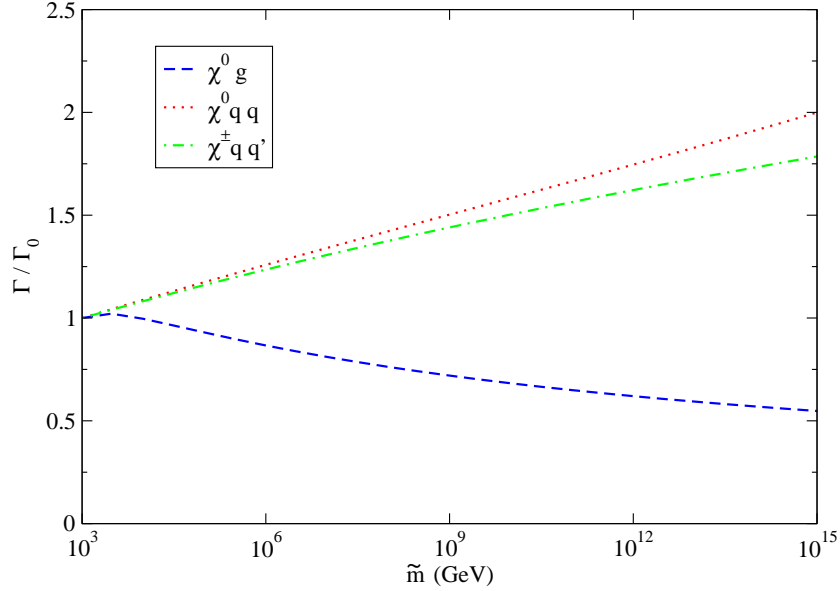}}
\end{center}
\caption{\sf Effect of the radiative corrections on the partial widths
for the decays $\tilde{g}\rightarrow\chi^0 g$ (dashed lines),
$\tilde{g}\rightarrow\chi^0 q \bar{q}$ (dotted) and
$\tilde{g}\rightarrow\chi^{\pm} q \bar{q}^{\,\prime}$ (dot--dashed) as
a function of $\msusy$.  The parameters are chosen as $\mgl = 1$ TeV,
$\tan\beta = 20$ and $\mu>0$.}
\label{fig:corrs}
\end{figure}

\begin{figure}[t]
\begin{center}
\mbox{\epsfig{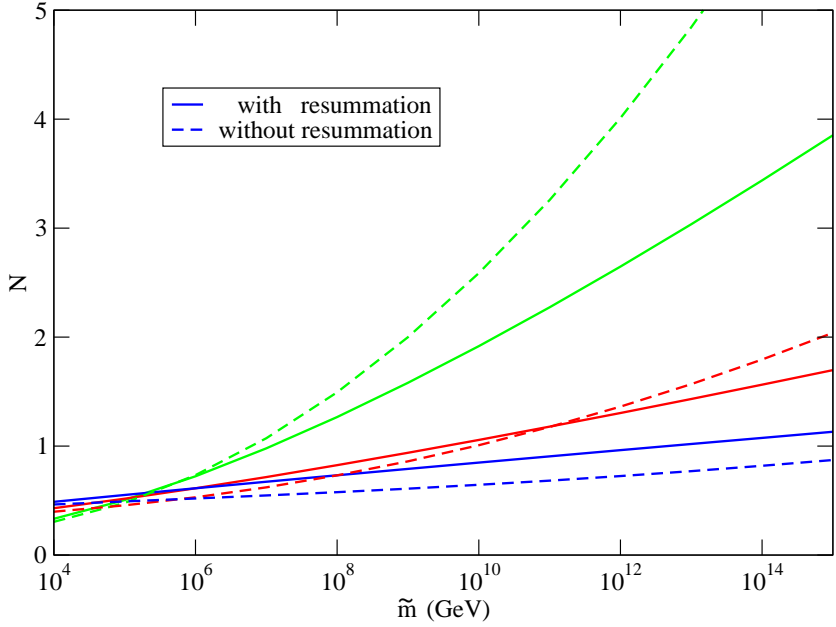}}
\end{center}
\vspace{-2mm}
\caption{\sf The normalization  $N$ of eq.~(\ref{scaling}) as
a function of the sfermion mass scale $\msusy$, with (solid lines) and
without (dashed lines) resummation of the leading logarithmic
corrections.  The upper, middle and lower sets of curves correspond to
$\mgl =$ 0.5, 1 and 2 TeV, respectively. The other free parameters are
chosen as $\tan\beta = 20$ and $\mu>0$.}
\label{fig:gamma}
\end{figure}

To further illustrate the effect of the resummation of the leading
logarithmic corrections, we plot in fig.~\ref{fig:corrs} the ratio
$\Gamma/\Gamma_0$ of the partial decay widths with and without
resummation, for the processes $\tilde{g} \rightarrow \chi^0 g$,
$\tilde{g} \rightarrow \chi^0 q \bar{q}$ and $\tilde{g} \rightarrow
\chi^{\pm} q \bar{q}^{\,\prime}$. We fix $\mgl = 1$ TeV, $\tan\beta =
20$ and $\mu>0$, but we have checked that the qualitative behaviour of the
corrections is independent of the precise choice of the parameters. It
can be seen from fig.~\ref{fig:corrs} that for large enough values of
$\msusy$ the radiative corrections can be of the order of 50--100\%,
and that they enhance the widths for the three--body decays and
suppress the width for the radiative decay.

To conclude this section, we discuss the scaling behaviour of the
gluino lifetime and total decay width. The lifetime $\tg =
\hbar/\gtot$ can be written as
\be
\label{scaling}
\tg = \frac{4 \,{\rm sec}}{N}\,
\times \,\left(\frac{\msusy}{10^9\,{\rm GeV}}\right)^4 \times 
\left(\frac{1 \,{\rm TeV}}{\mgl}\right)^5 ,
\ee
where the normalization $N$ is of order unity and depends on $\msusy$
and $\mgl$ (and only very mildly on $\tan\beta$). In \fig{fig:gamma}
we show $N$ as a function of $\msusy$ for $\tan\beta = 20$ and three
different values of the physical gluino mass ($\mgl = 0.5,\, 1$ and 2
TeV, respectively). The non--vanishing slope of $N$ represents the
deviation of the total gluino decay width from the naive scaling
behaviour $\gtot \propto \mgl^5\,/\,\msusy^4$. The solid lines in the
plot represent the results of our full calculation, whereas the dashed
lines represent the lowest--order results that do not include the
resummation. For low values of $\mgl$ the contribution of the
radiative decay dominates (see \fig{fig:bratios}), thus the total
decay width departs visibly from the naive scaling and is
significantly suppressed by the resummation of the radiative
corrections.  On the other hand, for large values of $\mgl$ the
three--body decays dominate, and the effect of the resummation is to
enhance the total decay width. Finally, for the intermediate value
$\mgl=1$ TeV there is a compensation between the corrections to the
radiative decay width and those to the three--body decay widths, and
the net effect on the total decay width of the resummation of the
leading logarithmic corrections is rather small.

\vspace{-1mm}
\section{Gluino Decay into Gravitinos} 
\label{sec5}  
\vspace{-1mm}

\sps\ opens up the possibility of direct tree-level mediation of the
original supersymmetry breaking to the SM superfields, without the
need of a hidden sector~\cite{noi4}. In usual low-energy
supersymmetry, this possibility is impracticable: for $F$--term
breaking some scalars remain lighter than the SM matter fermions, and
for $D$--term breaking gaugino masses cannot be generated at the same
order of scalar masses. In \sps\ a large hierarchy between scalar and
gaugino masses is acceptable, and indeed models have been
proposed~\cite{noi4,babu} with direct mediation of $D$--term
supersymmetry breaking.

Therefore, in \sps\ the original scale of supersymmetry breaking
$\sqrt{F}$, which is related to the gravitino mass by
\beq
m_{3/2}=\sqrt{\frac{8\pi}{3}} \frac{F}{\mpl},
\eeq
could be as low as the squark mass scale $\mtil$. This means that the
interactions between the gluino and (the spin--1/2 component of) the
gravitino, which are suppressed by $1/F$, could be as strong as those
considered in the previous sections, which are suppressed by
$1/\mtil^2$.

For $m_{3/2} \ll \mgl$, the gravitino interactions can be obtained,
through the supersymmetric analogue of the equivalence
theorem~\cite{equi}, from the goldstino derivative coupling to the
supercurrent. This approximation is valid as long as
$\sqrt{F} \ll 6\times (\mgl /1\,{\rm TeV})^{1/2} \times 10^{10}$~GeV. 
Using the equations of motion, we can write the
effective goldstino ($\wt G$) interactions for on--shell particles
as
\beq
{\cal L}=\frac 1F \left( -m_{{\tilde q}\lle}^2 {\tilde q}\lle {\bar q}\lle
-m_{{\tilde q}\rr}^2 {\tilde q}\rr {\bar q}\rr
+\frac{\mgl}{4\sqrt{2}}\,{\ov{{\tilde g}^a}}\, \sigma^{\mu\nu} 
\gamma_5 \,G^a_{\mu \nu}
\right) \tilde G +{\rm h.c.}
\label{goldlag}
\eeq
Below $\mtil$, the effective Lagrangian describing the interactions
between the gluino and the goldstino becomes
\beq
{\cal L} = \frac{1}{F}\,\sum_{i=1}^5 \CG_i\,\QG_i\,,
\label{eglag}
\eeq
\bea
\QG_1 & = &  \ov{\wt{G}} \,\gamma^{\mu}\,
\gamma_{5}\, {\tilde g}^a\; \otimes\;\sum_{k=1,2 \atop q=u,d}\;
\ov{q}^{\,(k)} \,\gamma_{\mu} \, T^a \,q^{\,(k)}
\label{opg1}\\
\QG_2 & = &  \ov{\wt{G}} \,\gamma^{\mu}\,
\gamma_{5}\, {\tilde g}^a\; \otimes\;
\ov{q}\lle^{\,(3)} \,\gamma_{\mu} \, T^a \,q\lle^{\,(3)}\\
\QG_3 & = &  \ov{\wt{G}} \,\gamma^{\mu}\,
\gamma_{5}\, {\tilde g}^a\; \otimes\;
\ov{t}\rr \,\gamma_{\mu} \, T^a \, t\rr \\
\QG_4 & = &  \ov{\wt{G}} \,\gamma^{\mu}\,
\gamma_{5}\, {\tilde g}^a\; \otimes\;
\ov{b}\rr \,\gamma_{\mu}\,  T^a\, b\rr \\
\QG_5 & = & \ov{\wt{G}} \,\sigma^{\mu\nu} \,\gamma_5\,
 {\tilde g}^a\;G^a_{\mu\nu}.
\label{opg5}
\eea
The Wilson coefficients at the matching scale $\mtil$ are
\be
\label{boundgoldstino}
\CG_1=\CG_2=\CG_3=\CG_4 = -\frac{g_s}{\sq2}\,,\;\;\;\;\;\;\;
\CG_5 = -\frac{\mgl}{2\,\sq2}.
\ee
Note that the coefficients of the interactions in \eq{eglag} have no
dependence on $\mtil$, because the squark mass square in the
propagators of the particles we integrate out is exactly cancelled by
the squark mass square in the goldstino coupling in \eq{goldlag}. 

The operator renormalization proceeds analogously to the discussion in
sect.~\ref{sec3}. The anomalous dimension matrix of the operators in
eqs.~(\ref{opg1})--(\ref{opg5}) is given by \eq{gamgam} with
\beq
\gamma_s^{\wt{\sss G}}=\frac 13 \pmatrix{
16-9N_c & 16 & 16 & 16 & 0 \cr
4 & 4-9N_c & 4 & 4 & 0 \cr
2 & 2 & 2-9N_c & 2 & 0 \cr
2 & 2 & 2 & 2-9N_c & 0 \cr
0 & 0 & 0 & 0 & 2N_f-18 N_c },
\eeq
\beq
\gamma_t^{\wt{\sss G}}= \pmatrix{
0 & 0 & 0 &  0 & 0 \cr
0 & 1 & \!\!\!\!-2 &  0 & 0 \cr
0 & \!\!\!\!-1 & 2 &  0 & 0 \cr
0 & 0 & 0 &  0 & 0 },~~~~\gamma_{st}^{\wt{\sss G}}=0\,.
\eeq

The evolution of the Wilson coefficients for the goldstino
operators has the simple analytic form
\bea
\CG_i (\mu) &=& \eta_s^{-\frac {9}{10}} \left[ \CG_i (\mtil ) +y \,{\ov
\CG}(\mtil )\right]
~~~i=1,4 \label{pincg}\\
\CG_2 (\mu) &=& \eta_s^{-\frac {9}{10}} \left[ (1+z)\,\CG_2 (\mtil )
-z \,\CG_3 (\mtil ) +y \,{\ov \CG}(\mtil )\right] ,\\
\CG_3 (\mu ) &=& \eta_s^{-\frac {9}{10}} \left[ (1+2\,z)\,\CG_3 (\mtil )
-2\,z \,\CG_2 (\mtil ) +y \,{\ov \CG}(\mtil )\right] \label{pallg},\\
\CG_5(\mu) &= &\eta_s^{-\frac {7}{5}}\,\CG_5 (\mtil )\,,
\eea
where ${\ov \CG}=2\,\CG_1 /3 +\CG_2/6+(\CG_3+\CG_4)/12$,
$y=\eta_s^{\,4/5}-1$, and $z=(\eta_s^{\,8/15}\,\eta_t^{-1/3}-1)/3$.  The
quantities $\eta_s$ and $\eta_t$ have been defined in
eqs.~(\ref{etas}) and (\ref{etat}), respectively.

\begin{figure}[t]
\begin{center}
\vspace*{-4mm}
\mbox{\epsfig{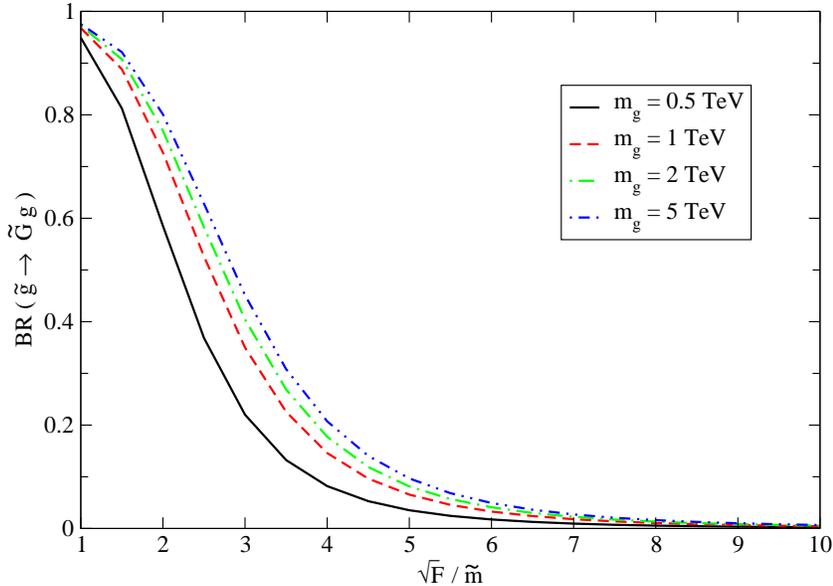}}
\end{center}
\vspace{-4mm}
\caption{\sf Branching ratio for the decay $\tilde g \rightarrow
\wt G \, g$ as a function of $\sqrt{F}/\msusy$, for different
values of the physical gluino mass $\mgl$. The other free parameters
are chosen as $\msusy = 10^{9}$ GeV, $\tan\beta = 2$ and $\mu>0$.}
\label{fig:goldstino}
\end{figure}

The formulae for the gluino decay widths into goldstino and quarks and
into goldstino and gluon can be found in the appendix. In
\fig{fig:goldstino} we show the branching ratio for the process
$\tilde g \rightarrow \wt G \, g$ as a function of the ratio
$\sqrt{F}/\msusy$, for $\msusy = 10^9$ GeV, $\tan\beta=2\,,\;\mu>0$
and different values of the gluino mass. The branching ratio for the
decay into goldstino and quarks, suppressed by phase space, is always
at or below the 1\% level. It can be seen from \fig{fig:goldstino}
that the gluino decay into goldstino and gluon is largely dominant
when $\sqrt{F}$ is as low as $\msusy$. In fact, the decays into
charginos or neutralinos and quarks (relevant for large values of
$\mgl$) are suppressed by phase space, while the radiative decay into
gluon and neutralinos (relevant for smaller values of $\mgl$) is
suppressed by $m_t^2/\mgl^2$ and a loop factor. With respect to the
scaling behaviour outlined in eq.~(\ref{scaling}), the additional
contribution to the total gluino decay width coming from the decay
into goldstino and gluon can significantly suppress the gluino
lifetime. In fact, for $\sqrt{F}= \msusy$ the normalization $N$ in
eq.~(\ref{scaling}) takes on values of order 40--50 for $\msusy >
10^8$ GeV.

On the other hand, the widths for the gluino decays into goldstino are
suppressed by a factor $\msusy^4/F^{2}$ with respect to those for
decays into charginos or neutralinos. Fig.~\ref{fig:goldstino} shows
that as soon as we depart from the condition $\sqrt{F}= \msusy$ the
branching ratio for $\tilde g \rightarrow \wt G \, g$ falls off very
quickly, and already for $\sqrt{F}/\msusy$ as large as 10 the gluino
decays into goldstino are below the 1\% level.

\vspace{-4mm}
\section{Conclusions}
\vspace{-4mm}
\label{concl}

If \sps\ is the correct theory to describe physics beyond the Standard
Model, one of its most spectacular manifestations might be the
discovery of a very long--lived gluino at the LHC. In this paper we
provided a precise determination of the gluino lifetime and branching
ratios. Applying to \sps\ the effective Lagrangian and renormalization
group techniques, we discussed the proper treatment of the radiative
corrections that are enhanced by the large logarithm of the ratio
between the sfermion mass scale and the gluino mass. We computed the
anomalous dimensions of the operators relevant to the gluino decay,
that allow us to resum to all orders in the perturbative expansion the
leading logarithmic corrections controlled by $\as$ and $\at$. We also
provided explicit analytical formulae for the gluino decay widths in
terms of the Wilson coefficients of the effective Lagrangian of
\sps. For representative values of the input parameters, we discussed
the numerical impact of the radiative corrections and found that they
can modify substantially the gluino decay width and branching ratios.
Finally, we considered models with direct mediation of supersymmetry
breaking, and we found that the gluino decays into gravitinos might
dominate over the other decay modes.

\newpage

\section*{Appendix}

We present in this appendix the explicit formulae for the leading
three--body and two--body gluino decay widths. All the results are
expressed in terms of the Wilson coefficients of the effective
Lagrangian of {\sps}, discussed in sects.~\ref{sec2}, \ref{sec3} and 
\ref{sec5}.

\paragraph{Three--body decays into quarks and chargino or neutralino:}
denoting the momenta of the decay products as $(p_1,p_2,p_3) \equiv
(p_{q_I},p_{\ov{q}_J},p_{\chi})$, and $s_{ij} = (p_i+p_j)^2$, the
three--body decay amplitude is given by
\be
\label{decay}
\Gamma_{\chi\, q_I\ov{q}_J} = \frac{1}{256\,\pi^3\,\mgl^3\;\mtil^4} \;\int \;
\ov{\left|{\cal M}\right|^2} \; ds_{13}\,ds_{23}\,.
\ee
The bar over $\left|{\cal M}\right|^2$ denotes the average over colour
and spin of the gluino and the sum over colour and spin of the final
state particles (the dependence on $\mtil$ has been factored out). The
limits of the integration in the ($\sut,\sdt$) plane are
\bea
\sut^{\rm max} &= &m_{q_I}^2 + \mch^2 + \frac{1}{2\,\sdt}\,
\left[(\mgl^2-m_{q_I}^2-\sdt)\,(\sdt-m_{\ov{q}_J}^2+\mch^2)\right.
\label{s13max}
\nn\\
&&\hspace{4cm}+\left.\lambda^{1/2}(\sdt,\mgl^2,m_{q_I}^2)
\,\lambda^{1/2}(\sdt,m_{\ov{q}_J}^2,\mch^2)\right]\,,\\
\sut^{\rm min} &= &m_{q_I}^2 + \mch^2 + \frac{1}{2\,\sdt}\,
\left[(\mgl^2-m_{q_I}^2-\sdt)\,(\sdt-m_{\ov{q}_J}^2+\mch^2)\right.
\nn\\
&&\hspace{4cm}-\left.\lambda^{1/2}(\sdt,\mgl^2,m_{q_I}^2)
\,\lambda^{1/2}(\sdt,m_{\ov{q}_J}^2,\mch^2)\right]\,,\\
\sdt^{\rm max} &= & (|\mgl|-m_{q_I})^2\,,\\
\sdt^{\rm min} &= & (|\mch|+m_{\ov{q}_J})^2\,,
\label{s23min}
\eea
where $\lambda(x,y,z) = x^2+y^2+z^2 -2\,(xy+xz+yz)$.

In the computation of the decays involving quarks of the first and
second generation we can neglect the quark masses and we find
\bea
\label{gammacharg}
\Gamma_{\chi_i^+d\ov{u}}=\Gamma_{\chi_i^-u\ov{d}}&=&
\frac{\mgl^5}{1536\,\pi^3\,\msusy^4}\left[
\left(\left.\CCL{1}\right.^2+\left.\CCR{1}\right.^2\right)\,g(x_i)
- 2\,\CCL{1}\,\CCR{1}\, f(x_i)\right]\,,\\
&&\nn\\
\label{gammaneut}
\Gamma_{\chi_i^0q\ov{q}}&=&
\frac{\mgl^5}{768\,\pi^3\,\msusy^4}\;
\left(\left.\CNL{1}\right.^2+\left.\CNR{1}\right.^2\,\right)\,
\biggr[g(x_i) + f(x_i)\biggr]\;\;\;\; (q=u,d)\,,
\eea
where $x_i = m_{\chi_i}/\mgl$, and we have included an overall factor 2 to
take into account the sum over the two generations of light
quarks. The functions $f$ and $g$ are defined as:
\bea
g(x) & = & 1-8\,x^2+8\,x^6-x^8-12\,x^4\,\ln x^2 \,,\\
f(x) & = & 2\,x+18\,x^3-18\,x^5-2\,x^7+12\,x^3(1+x^2)\,\ln x^2\,.
\eea

For generic quark masses the integration of the squared amplitude
$\ov{\left|{\cal M}\right|^2}$ on the $(\sut,\sdt)$ plane cannot be
performed analytically, and in order to compute the total decay width
we must resort to a numerical integration. 

The squared amplitude for the processes $\tilde{g}\rightarrow
\chi_i^+\,b\,\ov{t}$ and $\tilde{g}\rightarrow \chi_i^-\,t\,\ov{b}$ is
given by

\bea
\label{Mtb}
\ov{\left|{\cal M}\right|^2} &=& 
\left.\CCL{2}\right.^2\,(\mgl^2+m_t^2-\sut)\,(\sut-\mcpi^2-m_b^2)\nn\\
&& \nn\\
&+&\left.\CCR{2}\right.^2 \,(\mgl^2+m_b^2-\sdt)\,(\sdt-\mcpi^2-m_t^2) 
\nn\\
&& \nn\\
&+& \frac{1}{4}\,
\left(\left.\CCL{3}\right.^2+\left.\CCR{3}\right.^2\right)\;
(\mcpi^2+\mgl^2-\sut-\sdt)\,(\sut+\sdt-m_t^2-m_b^2)\nn\\
&& \nn\\
&+& 4\,
\left(\left.\CCL{4}\right.^2+\left.\CCR{4}\right.^2\right)\;
\biggr[(\mcpi^2+\mgl^2-\sut-\sdt)\,
(\sut+\sdt-m_t^2-m_b^2-4\,\mcpi^2)\nn\\
&&\hspace{4cm}+4\,(\sut-\mcpi^2)\,
(\sdt-\mcpi^2)-4\,m_t^2\,m_b^2\biggr]\nn\\
&& \nn\\
&+&2\,\CCL{2}\,\CCR{2}\,\mgl\,\mcpi\,
(\sut+\sdt-\mcpi^2-\mgl^2)\nn\\
&& \nn\\
&+&\left(\CCR{2}\,\CCR{3} + 12\,\CCR{2}\,\CCR{4} \right)
\,\mcpi\,m_t\,(\sdt-m_b^2-\mgl^2)\nn\\
&& \nn\\
&+&\left(\CCR{2}\,\CCL{3} + 12\,\CCR{2}\,\CCL{4} \right)
\,\mgl\,m_b\,(\sdt-m_t^2-\mcpi^2)\nn\\
&& \nn\\
&-&\left(\CCL{2}\,\CCR{3} - 12\,\CCL{2}\,\CCR{4} \right)
\,\mgl\,m_t\,(\sut-m_b^2-\mcpi^2)\nn\\
&& \nn\\
&-&\left(\CCL{2}\,\CCL{3} - 12\,\CCL{2}\,\CCL{4} \right)
\,\mcpi\,m_b\,(\sut-m_t^2-\mgl^2)\nn\\
&& \nn\\
&+&2\,\left(\CCL{3}\,\CCL{4} + \CCR{3}\,\CCR{4} \right)
\,\left[(\mgl^2+\mcpi^2-\sut-\sdt)\,(\sdt-\sut+m_b^2-m_t^2)\right.\nn\\
&&\hspace{4cm}
\left.+ 2\,m_b^2\,(\sdt-m_t^2-\mcpi^2)-2\,m_t^2\,(\sut-m_b^2-\mcpi^2)\right]\nn\\
&& \nn\\
&-&2\,\left(\CCL{3}\,\CCR{3} + 48\,\CCL{4}\,\CCR{4} \right)
\,\mgl\,\mcpi\,m_t\,m_b\;.
\eea

\newpage

The squared amplitude for the processes $\tilde{g}\rightarrow
\chi_i^0\,t\,\ov{t}$ and $\tilde{g}\rightarrow \chi_i^0\,b\,\ov{b}$ is
given by

\bea
\label{Mqq}
\ov{\left|{\cal M}\right|^2} &=&
\left(\left.\CNL{2}\right.^2+\left.\CNR{2}\right.^2\,\right)\;
\left[(\mgl^2+m_q^2-\sut)\,(\sut-m_q^2-\mczi^2)\right.\nn\\
&&\hspace{1cm}\left.+\,(\mgl^2+m_q^2-\sdt)\,(\sdt-m_q^2-\mczi^2)
+ 2\,\mgl\,\mczi\,(\mgl^2+\mczi^2-\sut-\sdt)\right]\nn\\
&& \nn\\
&+& \frac{1}{4}\,\left(\left.\CNL{3}\right.^2+\left.\CNR{3}\right.^2\,\right)
\,(\mczi^2+\mgl^2-\sut-\sdt)(\sut+\sdt-2\,m_q^2)\nn\\
&& \nn\\
&+&4\,\left(\left.\CNL{4}\right.^2+\left.\CNR{4}\right.^2\,\right)
\,\left[(\mczi^2+\mgl^2+4\,m_q^2-\sut-\sdt)
(\sut+\sdt-2\,m_q^2-4\,\mczi^2)\right.\nn\\
&& \hspace{4cm}\left.+4\,(\sut-m_q^2-\mczi^2)\,
(\sdt-m_q^2-\mczi^2)+8\,m_q^2\,\mczi^2)\right]\nn\\
&& \nn\\
&+&4\,\CNL{2}\,\CNR{2}\,m_q^2\,
(\sut+\sdt+4\,\mgl\,\mczi-2\,m_q^2)\nn\\
&& \nn\\
&+&\left(\CNR{2}\,\CNR{3}-\CNL{2}\,\CNL{3}\right)
\,m_q\,\left[\mczi\,(\mgl^2+m_q^2-\sut)+\mgl\,(\mczi^2+m_q^2-\sdt)\right]\nn\\
&& \nn\\
&+&\left(\CNR{2}\,\CNL{3}-\CNL{2}\,\CNR{3}\right)
\,m_q\,\left[\mczi\,(\mgl^2+m_q^2-\sdt)+\mgl\,(\mczi^2+m_q^2-\sut)\right]\nn\\
&& \nn\\
&+& 12\,\left(\CNR{2}\,\CNR{4}-\CNL{2}\,\CNL{4}\right)
\,m_q\,\left[\mgl\,(\mczi^2+m_q^2-\sdt)-\mczi\,(\mgl^2+m_q^2-\sut)\right]\nn\\
&& \nn\\
&+& 12\,\left(\CNR{2}\,\CNL{4}-\CNL{2}\,\CNR{4}\right)
\,m_q\,\left[\mczi\,(\mgl^2+m_q^2-\sut)-\mgl\,(\mczi^2+m_q^2-\sdt)\right]\nn\\
&& \nn\\
&+&2\,\left(\CNL{3}\,\CNL{4}+\CNR{3}\,\CNR{4}\right)
\,(\mgl^2+\mczi^2+2\,m_q^2-\sut-\sdt)\,(\sdt-\sut)\nn\\
&& \nn\\
&-&2\,\left(\CNL{3}\,\CNR{3}+48\,\CNL{4}\,\CNR{4}\right)
\,\mgl\,\mczi\,m_q^2\;\;\;\;\;\;\;\;\;\;\;\; (q=t,b).
\eea

We have checked that inserting in eqs.~(\ref{Mtb})--(\ref{Mqq}) the
high--energy (i.e.~non resummed) expressions for the Wilson
coefficients given in sects.~\ref{sec2} and \ref{sec3} we reproduce
the tree--level results of ref.~\cite{jim}. 

\newpage

\paragraph{Two--body decays into neutralino and gluon:}
the width for the radiative decay of the gluino, $\tilde{g}\rightarrow
g\,\chi_i^0$, is
\be
\Gamma_{\chi^0_i g} = 
\frac{(\mgl^2-\mczi^2)^3}{2\,\pi\,\mgl^3\,\msusy^4}\,
\left(\left.C^{\,{\chi}^0_i}_g\right._{\rm \!\!\!eff} \right)^{2}\,.
\ee
The use of the effective coefficient 
$\left.C^{\,{\chi}^0_i}_g\right._{\rm \!\!\!eff}$
defined in eq.~(\ref{coeffg}) allows us to reproduce the 
complete one--loop result when the resummation is switched off.

\paragraph{Decays into goldstino:}

the decay width into goldstino and quarks of the first and second generation
is:
\be
\Gamma_{\wt{G}\ov{q}q} =
\frac{\mgl^5}{192\,\pi^3\,F^2}\;\left.\CG_1\right.^{\,2}\,,
\ee
where we have summed over all four light quark flavours. 

The gluino decay width into goldstino and third--generation quarks is
as in eq.~(\ref{decay}), with $\mtil^4$ replaced by $F^2$. The squared
decay amplitude, which has to be integrated numerically on the
$(\sut,\sdt)$ plane, is given by
\bea
\ov{\left|{\cal M}\right|^2} &=&
\left(\left.\CG_{q\lle}\right.^2+\left.\CG_{q\rr}\right.^2\right)\;
\left[(\mgl^2+m_q^2-\sut)\,(\sut-m_q^2)
+\,(\mgl^2+m_q^2-\sdt)\,(\sdt-m_q^2)\right]\nn\\
&& \nn\\
&+&4\,\CG_{q\lle}\,\CG_{q\rr}\,m_q^2\,
(\sut+\sdt-2\,m_q^2)\;\;\;\;\;\;\;\;\;\;\;\;\; (q=t,b)\,,
\eea
where
\be
\CG_{t\lle} = \CG_{b\lle} = \CG_2\,,\;\;\;\;\;\;\;
\CG_{t\rr} = \CG_3\,,\;\;\;\;\;\;\;
\CG_{b\rr} = \CG_4\,.
\ee

Finally, the gluino decay width into gluon and goldstino is:
\be
\Gamma_{\wt{G} g} = 
\frac{\mgl^3}{2\,\pi\,F^2}\,\left.\CG_5\right.^{\,2}\,\;.
\ee

\section*{Acknowledgements}
We thank M.~Toharia and J.~Wells for precious help in the comparison
with the results of ref.~\cite{jim}. We also thank M.~Gorbahn,
U.~Haisch and P.~Richardson for useful discussions.  P.~S.~thanks the
CERN Theory Division and INFN, Sezione di Torino for hospitality
during the completion of this work.  The work of P.~G.\ is supported
in part by the EU grant MERG-CT-2004-511156 and by MIUR under contract
2004021808-009.

\newpage

%%%%%%%%%%%%%%%%%%%%%%%%%%%%%%%%%%%%%%%%%%%%%%%%%%%%%%%%%%%%%%%%%%%%%%%%

\end{document}